\documentclass[aps,prb,reprint,groupedaddress,superscriptaddress,longbibliography]{revtex4-2}
\usepackage{amsmath}
\usepackage{amssymb}
\usepackage{graphicx}
\usepackage{mathrsfs}
\usepackage{subfigure}
\usepackage{comment}
\usepackage{ntheorem}
\usepackage{tabularx}
\usepackage{color}
\usepackage{bbm}
\usepackage{dsfont}
\usepackage{bbold}
\usepackage[hidelinks]{hyperref}
\newcommand{\wsx}{\color {black}}
\newcommand{\zb}{\color {black}}
\newcommand{\zby}{\color {black}}

\begin{document}
\newtheorem{Definition}{Definition}[subsection]
   \title{Mechanism for scale-free skin effect in one-dimensional systems}
   \author{Shu-Xuan Wang}
   \email{shuxuanwang96@gmail.com}
   \affiliation{Guangdong Provincial Key Laboratory of Magnetoelectric Physics and Devices, State Key Laboratory of Optoelectronic Materials and Technologies, School of Physics, Sun Yat-sen University, Guangzhou 510275, China}
   \author{Zhongbo Yan}
   \email{yanzhb5@mail.sysu.edu.cn}
   \affiliation{Guangdong Provincial Key Laboratory of Magnetoelectric Physics and Devices, State Key Laboratory of Optoelectronic Materials and Technologies, School of Physics, Sun Yat-sen University, Guangzhou 510275, China}

   \date{\today}

   \begin{abstract}
        Non-Hermitian skin effect is one of the most captivating phenomena in non-Hermitian systems, {\zb characterized by the extensive localization of eigenstates near open boundaries. In its conventional form, the localization length of a skin mode is independent of the system size. Remarkably, however, when the open boundaries are coupled to realize a generalized boundary condition, the localization length can undergo a drastic transformation, becoming proportional to the system length. This intriguing regime is known as the scale-free skin effect (SFSE).
        Although SFSE has been observed in numerous one-dimensional non-Hermitian models through case-by-case studies, a unified theoretical framework capable of predicting its emergence {\zby and characteristic properties} remains absent. In this work, {\zby we take a firm step forward by establishing a model-independent framework for the analytic determination of scale-free localization lengths in certain regimes. Our key insight is to} treat generalized boundary conditions as a perturbation of the periodic boundary conditions, thereby circumventing the singular response-to-perturbation that would otherwise arise if they were perturbatively treated relative to open boundary conditions. Our work sheds new light on understanding SFSE in non-Hermitian systems.}
   \end{abstract}
         
   \maketitle

   \section{Introduction}
     Non-Hermitian physics has attracted tremendous interest over the past decade~\cite{RevModPhys.93.015005}. While conventional systems are described by Hermitian Hamiltonians, open systems—which exchange energy or particles with their environment—can be effectively modeled using non-Hermitian {\zb Hamiltonians}~\cite{PhysRevLett.70.2273,Rotter_2009,PhysRevLett.123.170401}.  In contrast to Hermitian systems, the energy spectrum of a non-Hermitian system under open boundary conditions (OBC) can differ drastically from its spectrum under periodic boundary conditions (PBC). 
     {\zb Moreover, the bulk eigenstates can all accumulate near the boundaries, a phenomenon known as the non-Hermitian skin effect (NHSE)}~\cite{PhysRevLett.121.086803,PhysRevLett.123.170401,PhysRevLett.123.016805,li2020critical,PhysRevResearch.2.043167,PhysRevB.104.195102,PhysRevB.106.085427,PhysRevA.106.062206,gu2022transient,manna2023inner,PhysRevB.108.L100301,PhysRevB.108.L220301,PhysRevB.107.155430,Li_2024,guo2024scale,PhysRevLett.133.216601,PhysRevB.106.L241112,PhysRevLett.133.076502,PhysRevLett.133.136503,PhysRevLett.133.136502,qin2025manybodycritical,wei2025generalized,shen2024nonhermitianhyperbolic,PhysRevB.111.035144,PhysRevB.111.115415,PhysRevLett.124.086801,PhysRevB.102.205118,PhysRevB.104.L161106,lei2024activating,vxgf-59xt,vz12-2qn3,PhysRevResearch.6.013213,PhysRevLett.131.116601,PhysRevB.103.045420}. To accurately describe the {\zb spectra} and eigenstates of such systems, non-Bloch band theory {\zb has been developed}~\cite{PhysRevLett.121.086803,PhysRevLett.123.066404,li2025phasespace,PhysRevB.99.201103,
     PhysRevLett.125.226402,PhysRevB.103.L241408,PhysRevB.107.L220301,PhysRevLett.131.076401,
     xiong2024skineffectarbitrary,PhysRevX.14.021011,wang2025generalgeometrydependent,cwwd-bclc,
     yi2025anomalousscaling,PhysRevResearch.5.043073,song2024fragilenonblochspectrum,
     wang2024nonblochselfenergy,PhysRevB.107.115412}. {\zb The NHSE has since been demonstrated and realized across a wide range of platforms}, including acoustic systems~\cite{zhang2021acoustic,xr3h-j9pv}, electric circuit~\cite{guo2024scale,PhysRevResearch.2.023265,zou2021observation,PhysRevB.107.085426,PhysRevResearch.2.022062,zhu2023higher,48wx-5gmj}, 
     photonic crystals~\cite{zhou2023observation,PhysRevLett.133.073803,PhysRevApplied.14.064076,2024_Wang,fang2022geometry}, 
     cold-atom systems~\cite{PhysRevLett.129.070401,zhao2025two},  and quantum dynamic systems~\cite{PhysRevResearch.2.043167,Wang_2023,li2024observation}.
     
    {\zb Within the framework of non-Bloch band theory}, the {\zb conventional Bloch} wave vector $k$  is generalized to a complex number $\tilde{k} = k + i \mu (k)$ for {\zb one-dimensional} non-Hermitian systems. This complex wave vector defines a parameter $\beta = e^{i \tilde{k} }$, whose values trace out one or more closed curves in the complex plane---referred to as the generalized Brillouin zone (GBZ) \cite{PhysRevLett.121.086803,PhysRevLett.123.066404,PhysRevLett.125.226402}. {\zb In close analogy to conventional Bloch theory, 
    the spectrum under OBC is determined by the non-Bloch Hamiltonian $H (\beta)$ evaluated on the GBZ, and the eigenstate corresponding to a given $\beta$ exhibits a spatial profile of the form $\psi (x)\sim\beta^x$. When $|\beta|<1$ or $|\beta|>1$,  the eigenstate decays or grows exponentially with $x$, respectively, corresponding to skin modes localized at the left or right boundary. Only when $|\beta|=1$ does the eigenstate become extended, resembling a conventional Bloch state.}

     {\zb The localization length of a skin mode is characterized by $\xi=|\ln \beta|^{-1}$. Typically, since $\beta$ is determined solely by the system parameters and is independent of the system size, the localization length is likewise size-independent. Such size independence is a common feature of wave-function localization, as exemplified by Anderson localization and topological boundary states in Hermitian systems. 
     Remarkably, recent studies on one-dimensional non-Hermitian systems have revealed that the localization length of skin modes can undergo a dramatic change when the two ends are coupled to implement a generalized boundary condition (GBC). In particular, it has been found that such a modification of the boundary conditions can shift the localization length from being size-independent to exhibiting a linear scaling 
     $\xi\sim L$ (with $L$ the system size). This anomalous scaling gives rise to the concept known as the scale-free skin effect (SFSE)~\cite{li2021impurity,PhysRevLett.127.116801,PhysRevB.110.214206,PhysRevB.109.L140102,Yuce_2025,PhysRevB.111.L140201,nnpf-8pxq,PhysRevResearch.5.033058}. Notably, although this phenomenon was first identified in systems with uniform non-Hermiticity, it has since been found to also emerge in a single-band Hermitian chain with a purely imaginary impurity at the boundary~\cite{PhysRevB.108.L161409}. This finding indicates that the NHSE is not a prerequisite for the emergence of eigenstates exhibiting scale-free localization behavior. The appearance of the same phenomenon in fundamentally distinct models challenges a 
     unified explanation of its origin, and consequently hinders the development of an effective method for establishing 
     the conditions under which such an effect arises in a given system. Indeed, current theoretical approaches for identifying scale-free skin modes remain largely limited to case-by-case numerical analyses of specific models  under GBC. }
     
     {\zb In this work, we develop an analytical approach based on perturbation theory to {\zby understand the origin of scale-free modes and determine their localization lengths}. As noted earlier, modifying OBC to GBC in systems exhibiting NHSE represents a prototypical route to SFSE, so it would seem natural to treat the boundary coupling as a perturbation of the open-boundary system. However, owing to the accumulation of a vast number of skin modes near the boundaries, even an arbitrarily weak boundary coupling can dramatically reshape the GBZ and energy spectra of non-Hermitian systems with NHSE, especially in the thermodynamic limit ~\cite{PhysRevB.109.144203,PhysRevResearch.7.L032043}. 
     Since $\beta$ dictates the localization length, its singular response to perturbations indicates that 
     the scale-free modes observed under GBC are not a simple perturbed variation of the skin mode under OBC. Therefore, 
     treating the GBC as a perturbation of OBC is not a viable approach. 
     
     We note that the spatial profile of a scale-free mode follows $\psi(x)\sim\beta^{x/L}$, which becomes extended 
     in the thermodynamic limit. Additionally, as $L\rightarrow\infty$, the spectrum under GBC converges to that under PBC. 
     This observation suggests that the properties of non-Hermitian systems under PBC are more stable against boundary 
     perturbations than those under OBC. Motivated by this key observation, we treat GBC as a perturbation of PBC, 
     thereby naturally circumventing the singular response that would arise if GBC were treated as a perturbation 
     of OBC. Notably, this method is model-independent 
     and {\wsx provide{\zby s} {\zby an analytical} framework to research SFSE}.}

     {\zb The remainder of this paper is organized as follows. In Sec.~\ref{II}, we establish the theoretical 
     framework. In Sec.~\ref{III}, we apply our theory to the Hatano–Nelson model to demonstrate its 
     predictive power. We then discuss the applicability and limitations of our approach in Sec.~\ref{IV}, 
     where we also present our concluding remarks. Detailed derivations of key formulas are provided 
     in the Appendices.}
     
    \section{Theoretical framework}\label{II}

      {\zb For generality, we start with a one-dimensional tight-binding Hamiltonian featuring 
      $w$ orbitals per unit cell. We restrict intercell hopping to nearest-neighbor terms only---a choice that entails no loss of generality, as any longer-range finite hopping can be recast into this form by an appropriate redefinition of the unit cell (i.e., by constructing a supercell). 
      The Hamiltonian is then written as}
        \begin{equation}
           \begin{split}
               H_{\rm{PBC}} = &\sum_{n=1}^{L-1} \hat{\psi}^{\dagger}_{n} h_{0} \hat{\psi}_{n} + \hat{\psi}^{\dagger}_{n+1} {\zb h_{1}} \hat{\psi}_{n} + \hat{\psi}^{\dagger}_{n} h_{-1} \hat{\psi}_{n+1} \\
                &+ \hat{\psi}^{\dagger}_{L} h_{0} \hat{\psi}_{L} + \hat{\psi}^{\dagger}_{1} {\zb h_{1}} \hat{\psi}_{L} + \hat{\psi}^{\dagger}_{L} h_{-1} \hat{\psi}_{1} ,
           \end{split}
           \label{1}
        \end{equation}
      where {\zb $h_n$ is a $w \times w$ matrix  encoding the intra-unit-cell couplings for $n=0$ and the inter-unit-cell hoppings for 
$n=\pm1$. The system has length $L$, and $\hat{\psi}_n = (\hat{c}_{n,1}, \hat{c}_{n,2}, \dots, \hat{c}_{n,w} )^T$  is the $w$-component annihilation operator at the $n$th unit cell, with its Hermitian conjugate $\hat{\psi}^{\dagger}_{n} = (\hat{c}^{\dagger}_{n,1}, \hat{c}^{\dagger}_{n,2}, \dots, \hat{c}^{\dagger}_{n,w} )$ representing the corresponding creation operator.}  

{\zb Following the non-Bloch band theory}, the corresponding non-Bloch Hamiltonian is given by
        \begin{equation}
            H (\beta) = h_0 + h_1 \beta^{-1} + h_{-1} \beta,
            \label{2}
        \end{equation}
      {\zb whose eigenvalues form $w$ bands, denoted as $E_{p} (\beta)$ for $p=1,2,\dots,w$.}  
      Under PBC, translational symmetry is preserved, and {\zb $\beta$ lies on the conventional} Brillouin zone (BZ):  
      $\beta_k = e^{ i k}$ with $k = \frac{2 \pi l}{L}$ (${\zb l}=1,2,\dots, L$). Accordingly, the spectra of the $w$ bands 
      are given by $E_{p} (k) = E_{p} (\beta_k)$.  The real-space wave function of $H_{\rm PBC}$ 
      corresponding to $E_{p} (k)$ is extended and can be expressed as 
        \begin{equation}
           | \Phi^{\rm R }_{p,k} \rangle =\frac{1}{\sqrt{L}} \sum_{n=1}^{L} \sum_{q=1}^{w} \beta_{k}^n \phi^{\rm R}_{p,k,q} \hat{c}_{n,q}^{\dagger} | 0 \rangle,
           \label{4}
        \end{equation}
      where the superscript ${\rm R}$ denotes right eigenstates, $q$ indexes the orbitals,  and $\phi^{\rm R}_{p,k} = (\phi^{\rm R}_{p,k,1}, \phi^{\rm R}_{p,k,2}, \dots, \phi^{\rm R}_{p,w})^T$ is the right eigenvector of $H(\beta_k)$ associated with the eigenvalue $E_{p} (k)$. Correspondingly, the wave function of the left eigenstate related to $| \Phi^{R }_{p,k} \rangle$ is
        \begin{equation}
              | \Phi^{\rm L}_{p,k} \rangle =\frac{1}{\sqrt{L}} \sum_{n=1}^{L} \sum_{q=1}^{w} \beta_{-k}^{-n} \phi^{\rm L}_{p,k,q} \hat{c}_{n,q}^{\dagger} | 0 \rangle,
              \label{5}
        \end{equation}
      where $\phi^{\rm L}_{p,k} = (\phi^{\rm L}_{p,k,1}, \phi^{\rm L}_{p,k,2}, \dots, \phi^{\rm L}_{p,w})^T$ is the left eigenvector of $H(\beta_k)$ {\zb that satisfies} $ ( \phi^{\rm L}_{p,k_1} )^{\dagger} \phi^{\rm R }_{q,k_2} = \delta_{p,q} \delta_{k_1,k_2} $.

      {\zb Next, we introduce a local term that acts on only a few unit cells. This term locally breaks 
      translational symmetry, and its position can therefore be regarded as a boundary under GBC. 
      Physically, such a term may be interpreted as a specific type of impurity. Its general form is given by}
        \begin{equation}
            H_{\rm imp} = \sum_{r,s \in boundary} \hat{\psi}^{\dagger}_{r} h_{r,s} \hat{\psi}_{s}.
            \label{6}
        \end{equation}
      The total Hamiltonian for the system is hence $H=H_{\rm PBC} + H_{\rm imp}$. Treating $H_{\rm imp}$ as a perturbation, $E_{p} (k)$ is modified as $ \tilde{E}_{p} (k)  \approx E_{p} (\beta_k) + \langle \Phi^{\rm L }_{p,k} | H_{\rm imp} | \Phi^{\rm R }_{p,k} \rangle $ \footnote{For general cases, the spectra of non-Hermitian systems are closed curves in the complex plane. Hence, $E_{p} (k)$ is typically non-degenerate except for some special cases. We therefore employ non-degenerate perturbation theory in this work.}. {\zb Since $H_{\rm imp}$ acts only locally near the boundary, 
      whereas both $ |\Phi^{\rm R }_{p,k}\rangle$ and $|\Phi^{\rm L }_{p,k}\rangle$ are uniformly delocalized with probability density 
      $1/L$ per unit cell, the eigenvalues corrected by the perturbation can be written as}
        \begin{equation}
              \tilde{E}_{p} (k)  \approx  E_{p} (\beta_k) + \frac{C_{p,k}}{L},
              \label{7}
        \end{equation}
      where $|C_{p,k}|$ is independent of $L$. {\wsx To ensure the accuracy of this approximation, we require that}  
        \begin{equation}
          \left| \frac{C_{p,k}}{ \beta_k \left. \frac{\partial E_{p} (\beta)}{\partial \beta} \right|_{\beta = \beta_k} }  \right| =\left| A_{p,k} + i B_{p,k} \right| < n_c {\wsx \ll 2\pi},
            \label{8}
        \end{equation}
      {\zb where {\zby $A_{p,k}$ and $B_{p,k}$ denote} the real and imaginary parts, respectively, of the complex quantity} $ C_{p,k} / (\beta_k  \frac{\partial E_{p} (\beta)}{\partial \beta} |_{\beta = \beta_k}) $, and {\zby $n_c$} is the upper bound of {\zby this} complex quantity over all $k$. The value of $n_c$ cannot be determined definitively because the accuracy of the eigenenergy correction decreases gradually as $n_c$ increases, and no critical value of $n_c$ exists at which the perturbation formula becomes abruptly invalid. According to the perturbation theory, the condition $n_c \ll 2 \pi$ is required  (see Appendix A for the derivation). In practice, {\zby we find that this method performs well 
      even for $n_c = 1$ (see numerical results in Sec.~\ref{III})}.

      On the other hand, {\zb the relationship between the eigenenergy and the decay factor of the 
      eigenstate is governed by the bulk Schrödinger equation, which remains independent of 
      boundary conditions. Obviously, the eigenenergy obtained via the perturbation approach must 
      be consistent with that derived from solving the bulk Schrödinger equation. 
      This leads to the relation}
      \begin{equation}
          \tilde{E}_{p} (k) = E_{p} (\tilde{\beta}_k) = E_{p} (\beta_k + \Delta \beta_k).
          \label{9}
        \end{equation}
        where $\tilde{\beta}_k=\beta_k + \Delta \beta_k$ denotes the decay factor modified by the perturbation.
        Expanding $E_{p} (\beta_k + \Delta \beta_k)$ and utilizing Eq.\eqref{7}, {\zb we arrive at the relation}
        \begin{equation}
           \left. \frac{\partial E_{p} (\beta)}{\partial \beta} \right|_{\beta = \beta_k} \Delta \beta_k \approx \frac{C_{p,k}}{L}.
           \label{10}
        \end{equation}
      Consequently, we have 
        \begin{equation}
            \tilde{\beta}_{k} = \beta_k + \Delta \beta_k = \beta_k \left( 1 + \frac{A_{p,k} + i B_{p,k}}{L}  \right),
            \label{11}
        \end{equation}
      and 
        \begin{equation}
             | \tilde{\beta}_k | ={\zb \left|1 + \frac{A_{p,k} + i B_{p,k}}{L}\right|}\approx \exp \left( \frac{A_{p,k}}{L} \right).
             \label{12}
        \end{equation}
      {\zb This is our central result. It implies that scale-free modes can naturally emerge 
      when the boundary condition is switched from PBC to GBC.} Interestingly, the direction of 
      localization is dictated by the sign of $A_{p,k}$. This result further suggests that 
      scale-free modes can originate from the modification of extended states, and the extended states
       can be viewed as a special class of scale-free modes with $A_{p,k}=0$.  

    \section{examples}\label{III}
      \subsection{Hatano-Nelson model with a boundary-coupling impurity}
        {\zb The one-band Hatano–Nelson (HN) model serves as a paradigmatic non-Hermitian platform, 
        exhibiting diverse phenomena such as the NHSE and SFSE. We thus employ this model to 
        illustrate the applicability of our theory.  
        The HN model under PBC takes the from  }
          \begin{equation}
            H_{\rm HN} = \sum_{n=1}^{L-1} t_r \hat{c}_{n+1}^{\dagger} \hat{c}_n + t_l \hat{c}_{n}^{\dagger} \hat{c}_{n+1} + t_r \hat{c}_{1}^{\dagger} \hat{c}_L + t_l \hat{c}_{L}^{\dagger} \hat{c}_{L},
            \label{13}
          \end{equation}
        {\zb where $t_{r}$ and $t_{l}$ denote right- and left-hopping amplitudes, respectively. 
        The impurity Hamiltonian that alters the boundary condition is given by}
          \begin{equation}
             {\zb H_{\rm imp}^{(c)}} = \mu_r t_r  \hat{c}_{1}^{\dagger} \hat{c}_{L} + \mu_l t_l  \hat{c}_{L}^{\dagger} \hat{c}_{1},
             \label{14}
          \end{equation}
        {\zb where the subscript $c$ indicates that this impurity term describes boundary coupling.}  For simplicity, we set $\mu_r = \mu_l= \mu $ and $t_r = m t_l$, where $\mu, m \in \mathbb{R}$. According to the theoretical framework established in Sec.~\ref{II}, the parameters $A^{(c)}_k$ and $B^{(c)}_k$  are obtained as
           \begin{equation}
             \begin{split}
                A^{(c)}_k (\mu,m)= \mu \frac{ 1-  m^2  }{1+ m^2 - 2 m \cos(2k)},
                \\
                B^{(c)}_k (\mu,m) = \mu \frac{-2 m  \sin (2k)}{1+ m^2 - 2 m \cos(2k)}.
             \end{split}
           \label{15}
        \end{equation}
      According to {\wsx the previous section, {\zby we set} $n_c = 1$}, so the allowed range of $\mu$ is approximately given by
        \begin{equation}
            \left|\mu \right| < \frac{(1-{\zb |m|})^2}{|1-m^2|}.
            \label{16}
        \end{equation}
      The demonstration details of  Eqs.~\eqref{15} and \eqref{16} are given in Appendix B.

      From Eq.~\eqref{15}, we find that $A^{(c)}_k (\mu,m) = - A^{(c)}_k (\mu,1/m)$. This implies that, for a fixed $\mu$, the direction of scale-free localization is governed by the ratio $t_r/t_l$, analogous to the NHSE in the HN model with OBC. For a given $m$, we have $A^{(c)}_k (0,m) = 0$ and $\text{sgn}(A^{(c)}_k (\mu,m))=-\text{sgn}(A^{(c)}_k (-\mu,m))$. Consequently, as $\mu$ is continuously tuned from a negative to a positive value, the corresponding scale-free mode becomes extended at $\mu=0$ and becomes localized at the opposite boundary as a scale-free mode for $\mu>0$. This behavior precisely corresponds to the scale-free to reversed scale-free transition discussed in Ref.~\cite{li2021impurity}.

      {\zb To explicitly demonstrate the usefulness of our theory within its regime of validity, we employ the 
      quantity known as the mean position to show the agreement between the results obtained from our 
      analytical theory and those from exact numerical calculations. The mean position is defined as
      \begin{equation}
      \frac{\langle x \rangle_{p,\beta}}{L} = \frac{1}{L} \frac{\langle \Phi^{\rm R }_{p} (\beta)| \hat{x} |\Phi^{\rm R }_{p} (\beta) \rangle}{\langle \Phi^{\rm R }_{p} (\beta)| \Phi^{\rm R }_{p} (\beta) \rangle}.
       \end{equation}
      In the thermodynamical limit $L\rightarrow\infty$, this dimensionless quantity equals 
      $1/2$ for an extended mode with $|\beta|=1$, and takes the values $0$ and $1$ for left- and right-skin modes, respectively. {\wsx For a scale-free mode, its value is finite and {\zby lies} in the range $(0,1)$ (see Appendix C for details).}}
 
      In Fig.~\ref{fig1}, we compare the numerical and analytical results for the case 
      $m=2$. According to Eq.~\eqref{16}, the perturbation approach is valid in the regime
      $|\mu|< \frac{1}{3}$. To visualize the mean positions of all eigenstates, we order the 
      eigenstates by the real parts of their eigenvalues in ascending order and use this index 
      as the transverse coordinate axis. In the figures, the red, orange, and green dots represent 
      the numerical results for system sizes $L = 75$, $100$, and $125$, respectively, while the blue 
      curve corresponds to the analytical result obtained in the thermodynamic limit.
      
      Figure~\ref{fig1a} presents the results for $\mu=0$. Since this case corresponds to the system under PBC, 
      all eigenstates are extended, and their mean positions are uniformly equal to $1/2$. The numerical results 
      for different system sizes agree remarkably well with the analytical prediction in the thermodynamic limit. 
      Figures~\ref{fig1b} and~\ref{fig1c} display the results for $\mu=-1/4$ and $\mu=1/4$, respectively. 
      In both cases, the numerical data show excellent agreement with the analytical results. This consistency, 
      together with the observed evolution of the mean positions as the system size increases, suggests that in 
      the thermodynamic limit, the mean positions of all eigenstates neither approach $0$ nor $1$, thereby 
      indicating that all eigenstates correspond to scale-free modes.

      Figures~\ref{fig1d} and \ref{fig1e} present the results for $\mu =-\frac{1}{2}$ and $\mu=1/2$, respectively. 
      {\wsx In these cases, we observe a 
      notable discrepancy between the numerical data and the analytical curves because the perturbation formula cannot give accurate corrections for eigenenergies under $|\mu|= 1/2$}. Nevertheless, the numerical results 
      indicate that all eigenstates remain scale-free modes. 
      {\wsx We conjecture that this is because, when {\zb \(\left| A_{p,k} + i B_{p,k}  \right| > 1\)}, the shift of $E_{p,k}$ is not well approximated by $C_{p,k}/L$, but remains proportional to $1/L$.  This leads to a modified decay factor, $\tilde{\beta}_k \approx \beta_k \exp \left( c / L \right)$, which supports 
      the conclusion that the eigenstates retain their scale-free character.}
      
      Figure~\ref{fig1f} presents the results for $\mu = -1$, corresponding to the system under OBC. 
      It is evident from the numerical data that all eigenstates become skin modes, with their wave functions localized at 
      the right boundary. In this case, our theory fails completely, which is unsurprising given that it is formulated on the basis of PBC.

      \begin{figure}     
          \centering
          \subfigure[]{\includegraphics[scale=0.42]{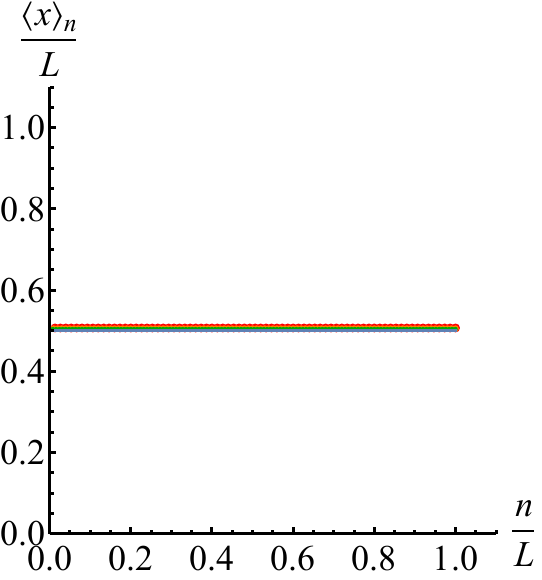}\label{fig1a}}
          \qquad
          \subfigure[]{\includegraphics[scale=0.42]{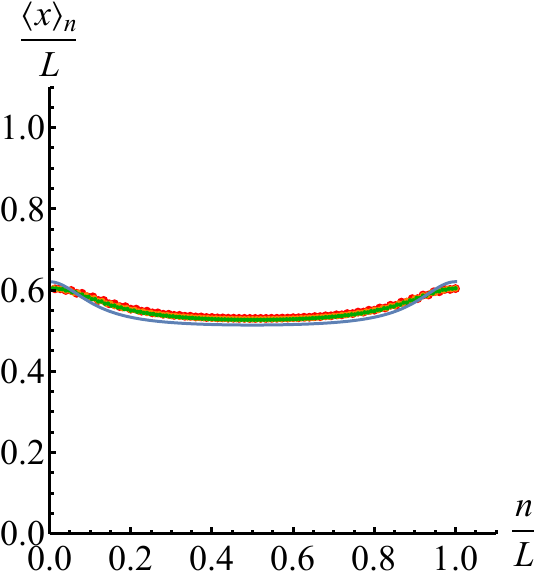}\label{fig1b}}
          \qquad
          \subfigure[]{\includegraphics[scale=0.42]{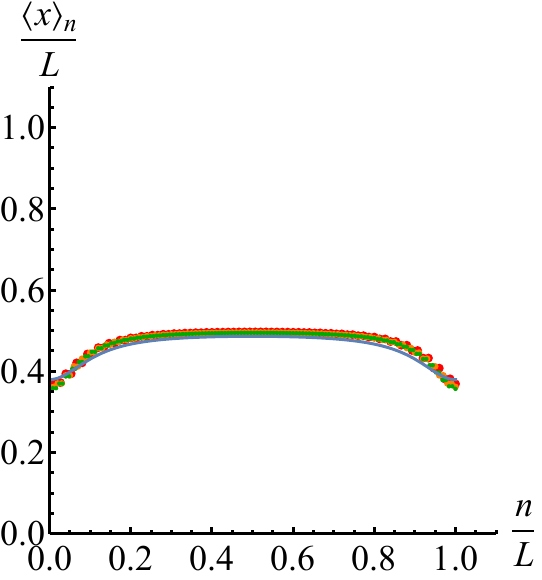}\label{fig1c}}
          \qquad
          \subfigure[]{\includegraphics[scale=0.42]{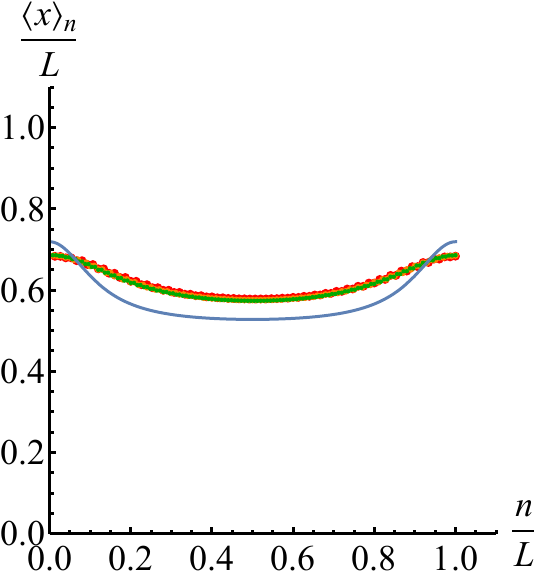}\label{fig1d}}
          \qquad
          \subfigure[]{\includegraphics[scale=0.42]{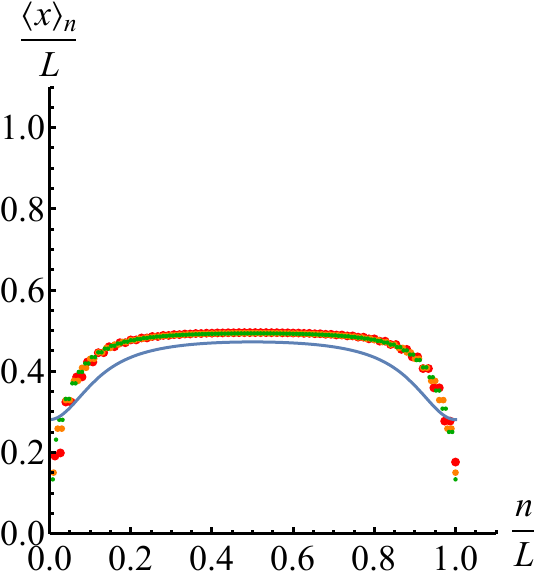}\label{fig1e}}
          \qquad
          \subfigure[]{\includegraphics[scale=0.42]{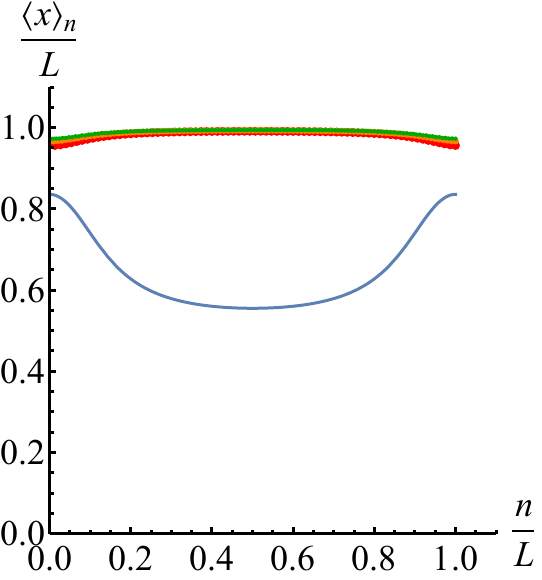}\label{fig1f}}
          \caption{Mean positions of the eigenstates of the Hatano–Nelson model with a boundary-coupling impurity. From panel (a) to panel (f), 
           the parameter $\mu$ takes the values $0$, $-\frac{1}{4}$, $\frac{1}{4}$, $-\frac{1}{2}$, $\frac{1}{2}$, and $-1$, respectively. The eigenstates are ordered by the index $n$, according to the ascending real parts of their eigenvalues. The red, orange, and green dots represent numerical results for system sizes $L=75$, $100$ and $125$, respectively, while the blue curves correspond to the theoretical predictions in the thermodynamic limit. The explicit expressions for these theoretical curves can be found in Appendix C. The remaining parameters are 
          $t_r = 2 t_l =2$.}
          \label{fig1} 
      \end{figure}

     \subsection{Hatano-Nelson model with an onsite boundary impurity}
     
     On-site potential represents another type of impurity. In this section, we extend our theoretical 
     framework to the HN model with a single on-site boundary impurity, described by the Hamiltonian
         \begin{equation}
           {\zb H_{\rm imp}^{(o)} }= V \hat{c}_1^{\dagger} \hat{c}_1,
           \label{17}
         \end{equation}
        where $V \in \mathbb{R}$. For this case, the parameters $A^{(o)}_k$ and $B^{(o)}_k$ are given by (see Appendix B for the derivation)
          \begin{equation}
             \begin{split}
               A^{(o)}_k  = {\wsx -} V \frac{ (t_l - t_r) \cos(k) }{t_l^2 + t_r^2 - 2 t_l t_r \cos(2k)},
              \\
              B^{(o)}_k  = - V \frac{(t_l + t_r) \sin(k)}{t_l^2 + t_r^2 - 2 t_l t_r \cos(2k)},
           \end{split}
           \label{18}
          \end{equation}
        The requirement {\wsx $n_c = 1$}  demands $V$ to satisfy the constrain:
          \begin{equation}
               |V| < |t_l - t_r|.
               \label{19}
          \end{equation}
        
        In the Hermitian case where \( t_l = t_r \), Eq.~\eqref{19} indicates that our framework is only valid for \( V = 0 \). This limitation arises because the Hermitian case exhibits energy degeneracies, which violates the fundamental assumption of non-degenerate perturbation theory upon which our framework is based. Consequently, our approach is invalid for Hermitian systems {\wsx under PBC}.
  
        In Fig.~\ref{2}, we set $t_r = 2 t_l = 2$ and compare the numerical results and analytical results for two distinct values of $V$.
        As in Fig.~\ref{2}(a), for $V = -\frac{1}{2}$, which satisfies the condition $|V|<|t_{r}-t_{l}|$, the numerical data 
        and the analytical curve are in excellent agreement, confirming that our perturbation method accurately captures the 
        spatial profile of the scale-free modes. Figure~\ref{2}(b) presents the case with $V=1$, corresponding to the critical situation. 
        The noticeable deviation observed for a subset of the skin-free modes indicates that the analytical theory 
        is approaching the limit of its validity.

        \begin{figure}     
          \centering
          \subfigure[]{\includegraphics[scale=0.42]{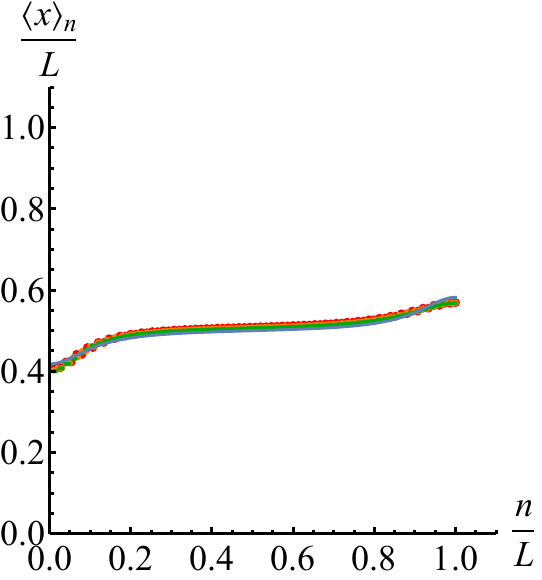}\label{fig2a}}
          \qquad
          \subfigure[]{\includegraphics[scale=0.42]{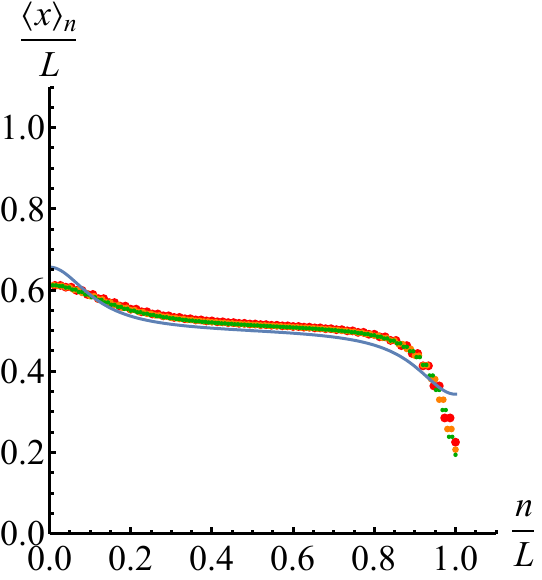}\label{fig2b}}
          \caption{Mean positions of the eigenstates of the Hatano–Nelson model with a single on-site boundary impurity, for (a) $V = -\frac{1}{2}$ 
          and (b) $V=1$. The eigenstates are ordered by the index $n$, according to the ascending real parts of their eigenvalues. The red, orange, and green dots represent numerical results for system sizes $L = 75$, $100$ and $125$, respectively, while the blue curves correspond to the theoretical predictions in the thermodynamic limit. The remaining parameters are $t_r = 2 t_l =2$.}
          \label{fig2} 
      \end{figure}

    \section{Discussion and Conclusion}\label{IV}
      In this work, we  have developed an analytical perturbation method to capture the localization behavior of the SFSE in one-dimensional non-Hermitian systems. Our framework treats systems with GBC as  PBC systems perturbed by certain types of boundary impurities. Since the wave functions under PBC are always extended, the eigenenergy shifts induced by these boundary impurities are proportional to the system size, which modifies the unitary decay factor into a scale-free one. We numerically verified our framework using the HN model with both coupling and onsite boundary impurities. In contrast to the case-by-case analyses in previous studies, {\zb our work provides a general and effective theoretical framework for determining the decay factor of skin-free modes induced by arbitrary types of local boundary perturbations, provided that the perturbation strength lies within the applicable regime.}

      The applicability of our framework is currently limited by the constraints of non-degenerate perturbation theory. Consequently, it cannot interpret all scale-free localization phenomena, such as the case of critical scale-free localization~\cite{PhysRevB.104.165117}. 
      {\zb Despite these limitations, our method remains broadly applicable and offers valuable insight by providing an analytical perspective on this distinctive non-Hermitian phenomenon.}

        {\wsx {\it Note added}.—Recently, the authors became aware of a related work \cite{li2026boundarysensitivenonhermiticityfloquethamiltonian}}

    \section{Acknowledgments}
          {\zb This work is supported by the Fundamental and Interdisciplinary Disciplines Breakthrough Plan of the Ministry of Education of China (Grant No. JYB2025XDXM403) and} the Guangdong Basic and Applied Basic Research Foundation (Grant No. 2023B1515040023).

    \appendix

    {\wsx
    \section{Derivation of Eq.\eqref{8}}
        We begin by reviewing the results of {\zby first-order} perturbation theory for non-degenerate systems. Consider a non-degenerate system described by the unperturbed Hamiltonian $H_0$, eigenstates $\{ |k_i^{(0)} \rangle \}$ ($i=1,2,3,\cdots$) and corresponding eigenenergies $\{E_i^{(0)} \}$. When a perturbation $H^{\prime}$ is introduced, the {\zby first-order} correction to the eigenenergy $E_i^{(0)}$ is given by
          \begin{equation}
            E_i = E_i^{(0)} + \langle k_i^{(0)} | H^{\prime} | k_i^{(0)} \rangle,
            \label{A1}
          \end{equation}
        while the corresponding eigenstate is modified as 
          \begin{equation}
               | k_i \rangle = |k_i^{(0)} \rangle + \sum_{j \not=i} \frac{\langle k_j^{(0)} | H^{\prime} | k_i^{(0)} \rangle}{E_i^{(0)} -E_j^{(0)}} |k_j^{(0)} \rangle.
               \label{A2}
          \end{equation}
        These expressions are accurate provided that, for all $i \not= j$, the off-diagonal matrix elements of $H^{\prime}$ satisfy
          \begin{equation}
              \left| \langle k_j^{(0)} | H^{\prime} | k_i^{(0)} \rangle  \right| \ll \left|  E_i^{(0)} -E_j^{(0)} \right|,
              \label{A3}
          \end{equation} 
        i.e., the perturbation couples different unperturbed levels only weakly compared to their energy gaps. Under this condition, each eigenstate receives only a small admixture from other unperturbed states via the perturbation.
        \par

        We now return to the system introduced in {\zby Sec.~\ref{II}}. For this system, the general condition in Eq.~(A3) takes the form
          \begin{equation}
             \left| \langle \Phi_{p,k}^L |H_{\rm imp}| \Phi_{p,k^{\prime}}^R \rangle \right| \ll \left| E_{p}(k) - E_{p} (k^{\prime}) \right|,
             \label{A4}
          \end{equation}
        for all $k^{\prime} \not= k$. Because the system has finite length $L$, for a fixed $k$, the minimum value of $\left| E_{p}(k) - E_{p} (k^{\prime}) \right|$ occurs at $k^{\prime} = k \pm 2\pi /L$. This minimum can be estimated as
          \begin{equation} 
           \begin{aligned}
              \min_{k^{\prime} \not= k}\left(\left| E_{p}(k) - E_{p} (k^{\prime}) \right|\right) 
              &= \left| E_{p}(k) - E_{p} (k \pm \frac{2 \pi}{L}) \right|   
               \\
              &\approx\frac{2 \pi}{L} \left|\frac{\partial E_p (k)}{\partial k}\right|
              \\
              &=\frac{2 \pi}{L} \left| \beta_k \left. \frac{\partial E_p (\beta)}{\partial \beta}\right|_{\beta=\beta_k}\right|.
           \end{aligned}
           \label{A5}
          \end{equation}
        In general, the diagonal matrix elements of the perturbation $H_{\rm imp}$ dominate over the off-diagonal ones; that is,
          \begin{equation}
              \left| \langle \Phi_{p,k}^L |H_{\rm imp}| \Phi_{p,k^{\prime}}^R \rangle \right| \leqslant \left| \langle \Phi_{p,k}^L |H_{\rm imp}| \Phi_{p,k}^R \rangle \right| = \left| \frac{C_{p,k}}{L} \right|,
              \label{A6}
          \end{equation}
        for all $k^{\prime} \not= k$. Consequently, for a fixed $k$, the condition Eq.~\eqref{A3} reduces to 
          \begin{equation}
              \left| \frac{C_{p,k}}{\beta_k \left. \frac{\partial E_p (\beta)}{\partial \beta}\right|_{\beta=\beta_k}} \right| \ll 2\pi.
              \label{A7}
          \end{equation}
        Let $n_c$ denote the upper bound of the quantity $ C_{p,k} / (\beta_k  \frac{\partial E_{p} (\beta)}{\partial \beta} |_{\beta = \beta_k}) $ over all $k$,  If we require the first-order corrections to be accurate for all eigenenergies, then according to Eq.~\eqref{A3} we need
          \begin{equation}
            n_c \ll 2\pi.
            \label{A8}
          \end{equation}
        Eqs.~\eqref{A7} and \eqref{A8} together yield Eq.~\eqref{8} of the main text.

    }

    \section{Derivation details for Eqs.\eqref{15}, \eqref{16}, \eqref{18} and \eqref{19}}
      The energy band for the single-band Hatano–Nelson model in Eq.~\eqref{13} is given by $E_{\rm HN} (\beta) = t_r \beta^ {-1} + t_l \beta$. Under PBC, we have $\beta \in \mathrm{BZ}$, i.e., $\beta = \beta_k = e^{ik}$, for $k=2 \pi l /L$ and $l = 1,2, \dots, L$. From Eqs.~\eqref{4} and \eqref{5},
      the right and left eigenstates corresponding to $E_{\rm HN} (\beta_k)$ are 
        \begin{equation}
          | \psi^{\rm R}_k \rangle = | \psi^{\rm L}_k \rangle =  \frac{1}{\sqrt{L}} \sum_{n=1}^{L} e^{i k n} \hat{c}_{n}^{\dagger} |0\rangle,
          \label{B1}
        \end{equation}
        Hence, the shift of $E_{\rm HN} (\beta_k)$ induced by $H_{\rm imp}^{(c)}$ is given by 
        \begin{equation}
             \Delta E_{k}^{(c)} = \langle \psi^{\rm L}_k | H_{c} | \psi^{\rm R}_k \rangle = \frac{1}{L} \left( \mu_r t_r e^{- i k} +  \mu_l t_l  e^{i k} \right).
             \label{B2}
        \end{equation}  
      Since
        \begin{equation}
          \beta_k \left. \frac{\partial E(\beta)}{\partial \beta} \right|_{\beta = \beta_k} = t_l \beta_k - t_r \beta_k^{-1},
          \label{B3}
        \end{equation}
      we have 
        \begin{equation}
               A^{(c)}_k + i B^{(c)}_k = \frac{\mu_r t_r e^{- i k} +  \mu_l t_l  e^{i k}}{t_l e^{i k} - t_r e^{-i k}}.
               \label{B4}
        \end{equation}
      Setting $\mu_r = \mu_l$ and $t_r = m t_l$, Eq.~\eqref{B4} reduces to 
        \begin{equation}
           \begin{split}
              A^{(c)}_k (\mu,m) = \mu \frac{ 1-  m^2  }{1+ m^2 - 2 m \cos(2k)},
              \\
              B^{(c)}_k (\mu,m) = \mu \frac{-2 m  \sin (2k)}{1+ m^2 - 2 m \cos(2k)}.
           \end{split}
           \label{B5}
        \end{equation}
      This corresponds to Eq.~\eqref{15} in the main text. Letting $|A^{(c)}_k|^2 + |B^{(c)}_k|^2 <{\zby n_{c}^{2}}$,  with $n_{c}=1$, we have 
        \begin{equation}
          \left| \mu \right|^2 <  \frac{\left( 1+ m^2 - 2 m \cos(2k)\right)^2}{ (1-m^2)^2 + 4 m^2 \sin^2(2k) }.
          \label{B6}
        \end{equation}
      {\zb For the convenience of discussion, we define} 
        \begin{equation}
            f (k) = \frac{ 1+ m^2 - 2 m \cos(2k)}{ \sqrt{(1-m^2)^2 + 4 m^2 \sin^2(2k)} }.
            \label{B7}
        \end{equation}
      {\zb The extrema of $f(k)$ are determined by the zeros of its derivative:}
        \begin{equation}
           \frac{\partial f}{\partial k} = \frac{4m \sin(2k) (1+m^2) (1+ m^2 - 2 m \cos(2k))}{(1+m^4 - 2 m^2 \cos(4k))^{3/2}} =0,
           \label{B8}
        \end{equation}
      {\zb It is straightforward to see that the solutions are $k=0$ and $\pm \frac{\pi}{2}$. Therefore, the extrema of $f(k)$ occur at these points. Evaluating $f(k)$ at these extrema gives}
        \begin{equation}
          f(0) = \frac{(1-m)^2}{|1-m^2|}, \quad f(\pm \frac{\pi}{2}) = \frac{(1+m)^2}{|1-m^2|},
          \label{B9}
        \end{equation}
      {\zb From these expressions, one finds that the minimum value of $f(k)$ is $(1-|m|)^2 / |1-m^2|$, regardless of the sign of $m$.} 
      If Eq.~\eqref{B6} holds for all $k$, we need
        \begin{equation}
              \left|\mu \right| < \frac{(1-{\wsx |m|})^2}{|1-m^2|},
        \end{equation} 
      which {\zb corresponds to Eq.~\eqref{16} in the main text.}

      {\zb For the case with a single on-site boundary impurity case, the} shift of $E_{\rm HN} (\beta_k)$ induced by $H_{\rm imp}^{(o)}$ is
      {\zb given by} 
        \begin{equation}
            \Delta E_{k}^{(o)} = \langle \psi^{\rm L}_k | H_{o} | \psi^{\rm R}_k \rangle = \frac{V}{L}.
            \label{B11} 
        \end{equation}
      Hence, {\zb we have} 
        \begin{equation}
              A^{(o)}_k + i B^{(o)}_k = \frac{V}{\beta_k \left. \frac{\partial E(\beta)}{\partial \beta} \right|_{\beta = \beta_k}} = \frac{V}{t_l e^{i k} - t_r e^{-i k}},
              \label{B12}
        \end{equation}
      and 
        \begin{equation}
           \begin{split}
               A^{(o)}_k  ={\wsx -} V \frac{ (t_l - t_r) \cos(k) }{t_l^2 + t_r^2 - 2 t_l t_r \cos(2k)},
              \\
              B^{(o)}_k  = - V \frac{(t_l + t_r) \sin(k)}{t_l^2 + t_r^2 - 2 t_l t_r \cos(2k)},
           \end{split}
           \label{B13}
        \end{equation}
      {\zb which corresponds to Eq.~\eqref{18} in the main text}. {\zb The condition} $|A^{(o)}_k + i B^{(o)}_k| < 1$ {\zb leads to 
      the constrain}
        \begin{equation}
            \frac{|V|^2}{t_l^2 + t_r^2 - 2 t_l t_r \cos(2k)} <1.
            \label{B14}
        \end{equation}
      {\zb For Eq.~\eqref{B14} to hold for all $k$, we need}
        \begin{equation}
            |V| < |t_l - t_r|,
            \label{B15}
        \end{equation}
      {\zb which corresponds to Eq.~\eqref{19} in the main text.}

    \section{Theoretical formula of the {\zb expectation} $\langle x \rangle/L$ under {\zb the} thermodynamic limit for skin and scale-free skin {\zb modes}}
        For {\zb an} eigenstate with decay factor $\beta$, the form of {\zb its} wave function is similar to Eq.~\eqref{4},
          \begin{equation}
              | \Phi^{\rm R }_{p} (\beta) \rangle =\frac{1}{\sqrt{L}} \sum_{n=1}^{L} \sum_{q=1}^{w} \beta^n \phi^{\rm R}_{p,\beta,q} \hat{c}_{n,q}^{\dagger} | 0 \rangle,
              \label{C1}
          \end{equation}
        where we set $ ( \phi^{\rm R}_{p,\beta} )^{\dagger} \phi^{\rm R }_{p,\beta}  =1$. The mean position is defined as the expectation value of the position operator with respect to the eigenstate $| \Phi^{\rm R }_{p} (\beta) \rangle$, namely,
          \begin{equation}
            \begin{split}
                \frac{\langle x \rangle_{p,\beta}}{L} &= \frac{1}{L} \frac{\langle \Phi^{\rm R }_{p} (\beta)| \hat{x} |\Phi^{\rm R }_{p} (\beta) \rangle}{\langle \Phi^{\rm R }_{p} (\beta)| \Phi^{\rm R }_{p} (\beta) \rangle}
                \\
                & = \frac{1}{L} \frac{\sum_{n=1}^{L} n |\beta|^2}{\sum_{n=1}^{L} |\beta|^2 }
                \\
                & = \frac{1- (1+L)|\beta|^{2 L} + L |\beta|^{2+2L}}{L (|\beta|^2 -1)(|\beta|^{2 L} -1)}.
            \end{split}
            \label{C2}
          \end{equation}
        For {\zb a} skin mode, $|\beta|>0$ is a constant. Thus, under {\zb the} thermodynamic limit, we have 
          \begin{equation}
              \lim_{L \rightarrow \infty} \frac{\langle x \rangle_{p,\beta}}{L} = 
                \begin{cases}
                  0 & \text{if} \ |\beta| < 1,
                  \\
                  \frac{1}{2} & \text{if} \ |\beta| = 1,
                  \\
                  1 & \text{if} \ |\beta| >1.
                \end{cases} 
                \label{C3}
          \end{equation}
        {\zb In contrast, for a scale-free mode, $\beta$ scales with the system size as $|\beta| = \exp(\frac{A}{L})$}. {\zb Accordingly}, under the thermodynamic limit, {\zb we have} 
          \begin{equation}
             \lim_{L \rightarrow \infty} \frac{\langle x \rangle_{p,|\beta|=\exp(A/L)}}{L} = \frac{1+(2 A -1) e^{2 A}}{ 2 A (e^{2A} -1)}.
             \label{C4}
          \end{equation}
          {\wsx
        For simplicity, denoting
          \begin{equation}
              g (A) =   \frac{1+(2 A -1) e^{2 A}}{ 2 A (e^{2A} -1)}
              \label{C5}
          \end{equation}
        For {\zb an} extended state, we have $A=0$, and 
          \begin{equation}
            \lim_{A \rightarrow 0}  g(A) = \frac{1}{2},
            \label{C6}
          \end{equation}
        which is consistent with {\zb the result shown in the middle line of Eq.~\eqref{C3}}. Let $A$ be {\zb given by} $A_{p,k}$ obtained for {\zb a} specific model,  then $g(A_{p,k})$ gives the theoretical curve for {\zb the} mean positions of the scale-free modes. Since
          \begin{equation}
            \lim_{A \rightarrow \infty } g(A) = 1, 
            \qquad 
            \lim_{A \rightarrow -\infty } g(A) = 0,
            \label{C7}
          \end{equation}
        the range for mean positions of the scale-free modes is $(0,1)$.
         }

   \bibliography{scale-free}

\end{document}